\documentclass[12pt,prb,aps,superscriptaddress,preprintnumbers,amsmath,amssymb]{revtex4-2}

\usepackage{graphicx} 
\usepackage{dcolumn} 

\def\bk{{\bf k}}

\def\bq{{\bf q}}

\def\bQ{{\bf Q}}

\def\b0{{\bf 0}}

\def\cO{{\cal O}}

\def\Re{{\rm Re}}
\def\Im{{\rm Im}}

\def\eps{\epsilon}

\def\Lam{\Lambda}
\def\om{\omega}

\def\Sg{\Sigma}

\def\tk{\tilde k}
\def\tq{\tilde q}

\begin{document}

\title{Non Fermi liquid behavior at flat hot spots from quantum critical fluctuations at the onset of charge- or spin-density wave order}

\author{Lukas Debbeler}
\affiliation{Max Planck Institute for Solid State Research,
 D-70569 Stuttgart, Germany}
\author{Walter Metzner}
\affiliation{Max Planck Institute for Solid State Research,
 D-70569 Stuttgart, Germany}

\date{\today}

\begin{abstract}
We analyze quantum fluctuation effects at the onset of charge or spin density wave order with a $2k_F$ wave vector $\bQ$ in two-dimensional metals -- for the special case where $\bQ$ connects a pair of hot spots situated at high symmetry points of the Fermi surface with a vanishing Fermi surface curvature. We compute the order parameter susceptibility and the fermion self-energy in one-loop approximation. The susceptibility has a pronounced peak at $\bQ$, and the self-energy displays non-Fermi liquid behavior at the hot spots, with a linear frequency dependence of its imaginary part. The real part of the one-loop self-energy exhibits logarithmic divergences with universal prefactors as a function of both frequency and momentum, which may be interpreted as perturbative signatures of power laws with universal anomalous dimensions. As a result, one obtains a non-Fermi liquid metal with a vanishing quasiparticle weight at the hot spots, and a renormalized dispersion relation with anomalous algebraic momentum dependencies near the hot spots.  
\end{abstract}

\maketitle


\section{Introduction}

Quantum fluctuations at and near quantum critical points (QCP) in metallic electron systems naturally lead to non-Fermi liquid behavior with unconventional temperature, momentum, and frequency dependencies of physical observables \cite{loehneysen07}. The fluctuation effects are most pronounced in low dimensional systems. In view of non-Fermi or ``strange metal'' behavior observed in various layered compounds, such as the high-$T_c$ cuprates, two-dimensional systems have attracted particular interest.
Due to the complex interplay of critical order parameter fluctuations with gapless electronic excitations, the theory of such systems is notoriously difficult.

Metals at the onset of charge or spin-density wave order provide a vast playground of quantum critical non-Fermi liquids with many distinct universality classes. The most intensively studied case of a N\'eel antiferromagnet is just one example \cite{abanov03,metlitski10_af1,lee18,berg12}.
A particularly intriguing situation arises when the wave vector $\bQ$ of the density wave is a {\em nesting vector}\/ of the Fermi surface, that is, when it connects Fermi points with collinear Fermi velocities \cite{footnote_perfnest}. Charge and spin susceptibilities exhibit a singularity at such wave vectors due to an enhanced phase space for low-energy particle-hole excitations. Since the nesting vectors in continuum systems are related to the Fermi momentum $k_F$ by the simple relation $|\bQ|=2k_F$, one may refer to such nesting vectors also as ``$2k_F$'' vectors \cite{altshuler95}.
The wave vector $(\pi,\pi)$ of a N\'eel state (in two dimensions) is a nesting vector only for special electron densities. Quantum critical fluctuations lead to an enhanced quasiparticle decay rate in this case, but not to a breakdown of Fermi liquid theory \cite{bergeron12,wang13}.

Recently, non-Fermi liquid behavior at the onset of charge- or spin-density wave order with {\em incommensurate}\/ \cite{fn_incomm} nesting wave vectors $\bQ$ in two-dimensional metals has been analyzed in a series of papers. 
In a one-loop calculation of the fermionic self-energy, a breakdown of Fermi liquid behavior was obtained at the {\em hot spots}\/ on the Fermi surface connected by the ordering wave vector \cite{holder14}. If the ordering wave vector $\bQ$ connects only a single pair of hot spots, in axial or diagonal direction, the frequency dependence of the one-loop self-energy at the hot spots obeys a power-law with exponent~$\frac{2}{3}$. If $\bQ$ connects two pairs of hot spots, the imaginary part of the real frequency one-loop self-energy exhibits a linear frequency dependence.
In none of these two cases the perturbative solution is self-consistent, and the feedback of the non-Fermi liquid self-energy seems to push the ordering wave vector away from the nesting point \cite{sykora18,sykora21}. Actually it was argued already long ago, for the case of a single hot spot pair, that quantum fluctuations spoil the QCP in favor of a first order transition \cite{altshuler95}.
However, a flattening of the Fermi surface at the hot spots might save the QCP \cite{sykora18}, and this scenario was supported by a systematic $\epsilon$-expansion around the critical dimension $\frac{5}{2}$ \cite{halbinger19}.
For the two hot-spot pair case, a self-consistent solution with a stable QCP was found numerically \cite{sykora21}.
While fluctuations are naturally stronger in two dimensional systems, quantum fluctuation effects at the onset of density wave order with a $2k_F$ wave vector are special and intriguing also in three dimensions \cite{schaefer17}.

\begin{figure}
\centering
\includegraphics[width=8cm]{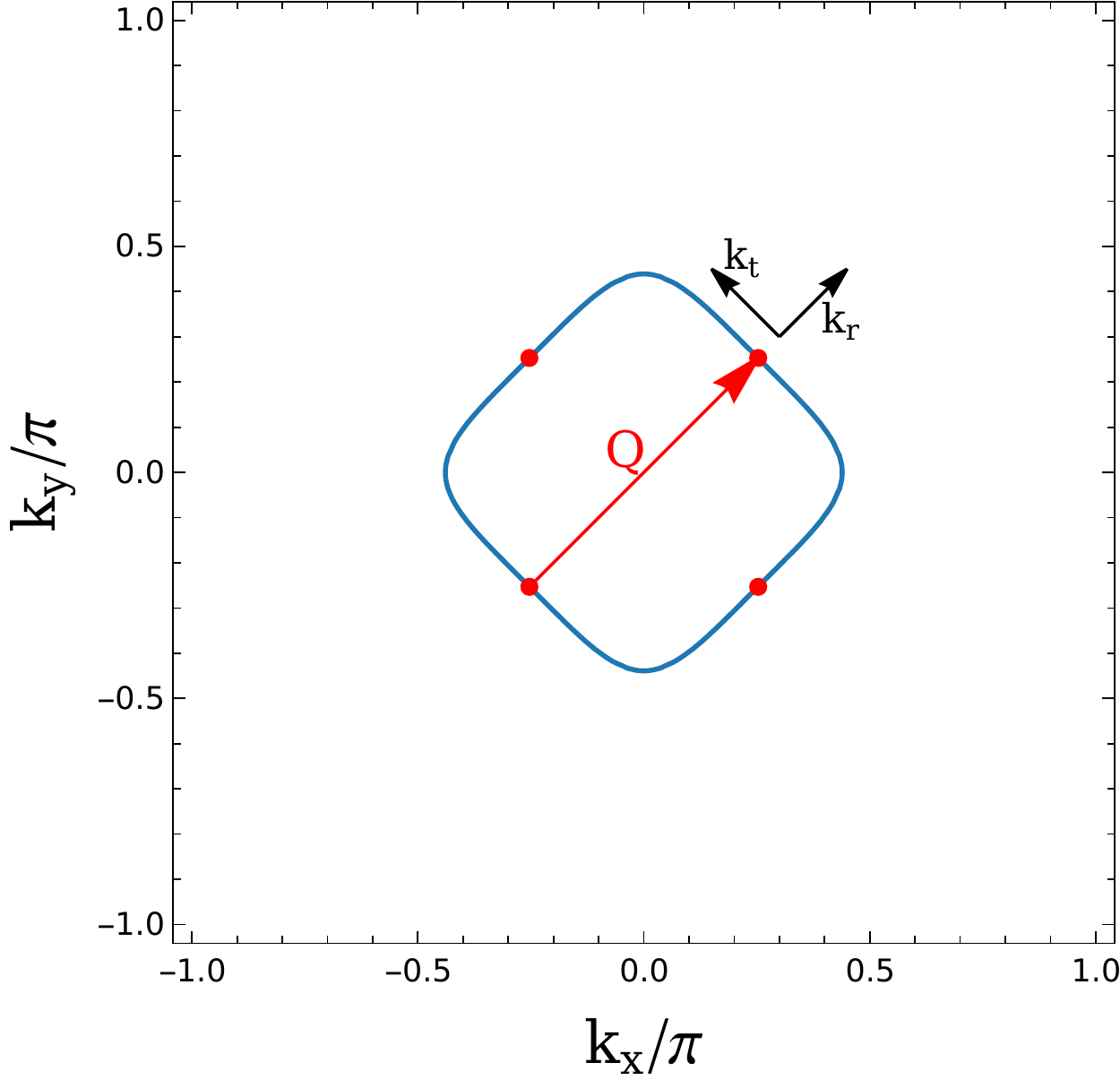}
\caption{Hot spots with vanishing curvature on the Fermi surface for a tight-binding model with nearest and next-nearest neighbor hopping amplitudes $t$ and $t'$, respectively, on a square lattice. The ratio of hopping amplitudes has been choosen as $t'/t = -0.35$, and the Fermi level leading to flat hot spots is $\eps_F = 8t'[1 - 2(t'/t)^2] = -2.114 t$.}
\end{figure}
In this paper we analyze quantum fluctuations and non-Fermi liquid behavior at the onset of density wave order in a two-dimensional system for a case where the nesting vector connects {\em flat}\/ hot spots on a mirror symmetry axis, where the Fermi surface curvature vanishes already in the non-interacting reference system, that is, before fluctuations are taken into account. Such a situation may arise at special electron filling factors. For example, for a tight-binding model with nearest and next-nearest neighbor hopping on a square lattice, the Fermi surface exhibits zero curvature points along the Brillouin zone diagonal for a specific choice of the Fermi level (corresponding to a special filling factor), as illustrated in Fig.~1.
Using relative momentum coordinates $k_r$ and $k_t$ in normal and tangential directions, respectively, with respect to the Fermi surface at a hot spot, the dispersion relation near the hot spot has the form
\begin{equation} \label{dispersion}
 \xi_\bk = \eps_\bk - \eps_F = v_F k_r + b k_t^4 \, ,
\end{equation}
to leading order in $k_r$ and $k_t$. Here $v_F$ is the Fermi velocity at the hot spot, and $b$ is a real constant, which is positive (negative) if the Fermi surface is convex (concave) at the hot spot. Due to the mirror symmetry with respect to the Brillouin zone diagonal there is no term of order $k_t^3$. Hence, this case differs from inflection points on the Fermi surface, where the curvature vanishes, too, but the leading tangential momentum dependence is of cubic order.

As in the above-mentioned case of N\'eel order with nested hot spots, a QCP with flat hot spots connected by an incommensurate wave vector requires tuning to a specific particle density. In addition, another parameter must be tuned such that the system is situated at the onset of charge or spin density wave order. In solids, the density of electrons in layered compounds can be varied over a broad range by chemical substitution or gate potentials. The tuning of other parameters such as the ratio of Coulomb to kinetic energy is in principle possible by pressure, but in practice limited to a narrow regime. Alternatively, a QCP with flat hot spots may be realized by cold fermionic atoms in optical lattices, where the particle density, interaction strength, and hopping amplitudes can be tuned at will \cite{gross17,hofstetter18}.

We compute the order parameter susceptibility, the effective interaction, and the fermion self-energy at the onset of incommensurate charge or spin density wave order with flat nested hot spots in a one-loop approximation. The susceptibility and the effective interaction exhibit pronounced peaks at the nesting vector. Both the momentum and frequency dependencies of the self-energy develop logarithmic divergencies, signalling non-Fermi liquid power-laws with universal critical exponents.

The remainder of the paper is structured as follows. In Sec.~II we compute the order parameter susceptibility and the effective interaction at the QCP. The momentum and frequency dependence of the fermion self-energy is evaluated in Sec.~III. A conclusion in Sec.~IV closes the presentation.


\section{Susceptibility and effective interaction}

We consider a one-band system of interacting fermions with a bare single-particle energy-momentum relation $\eps_\bk$. We are dealing exclusively with ground state properties, that is, the temperature is fixed to $T=0$. The bare fermion propagator has the form
\begin{equation}
 G_0(\bk,ik_0) = \frac{1}{ik_0 - \xi_\bk} \, ,
\end{equation}
where $k_0$ denotes the (continuous) imaginary frequency, and $\xi_\bk = \eps_\bk - \mu$. At zero temperature, the chemical potential $\mu$ is equal to the Fermi level $\eps_F$.
We assume that, in mean-field theory, the system undergoes a charge or spin-density wave transition with an incommensurate and nested wave vector $\bQ$, which connects a pair of hot spots on the Fermi surface. We further assume that the dispersion relation in the vicinity of the hot spots has a quartic tangential momentum dependence of the from Eq.~\eqref{dispersion}.

In random phase approximation (RPA, that is, at one-loop level), the order parameter susceptibility has the form
\begin{equation} \label{chi}
 \chi(\bq,iq_0) = \frac{\chi_0(\bq,iq_0)}{1 + g \chi_0(\bq,iq_0)} \, ,
\end{equation}
where $g<0$ is the coupling constant parametrizing the interaction in the instability channel.  The bare charge or spin susceptibility $\chi_0$ is related to the particle-hole bubble $\Pi_0$ by $\chi_0(\bq,iq_0) = - N \Pi_0(\bq,iq_0)$, where $N$ is the spin multiplicity, and \cite{negele87}
\begin{equation} \label{Pi0def}
 \Pi_0(\bq,iq_0) = \int_\bk \int_{k_0} G_0(\bk,ik_0) \, G_0(\bk-\bq,ik_0-iq_0) \, .
\end{equation}
Here and in the following $\int_\bk$ is a shorthand notation for $\int \frac{d^2\bk}{(2\pi)^2}$, and $\int_{k_0}$ for $\int \frac{dk_0}{2\pi}$.
While keeping the spin multiplicity $N$ as a general parameter in our equations, we choose $N=2$, corresponding to spin-$\frac{1}{2}$ fermions, in all numerical results.
Continuing $\Pi_0(\bq,iq_0)$ analytically to the real frequency axis from the upper complex frequency half-plane yields the retarded polarization function $\Pi_0(\bq,\om)$.
\begin{figure}
\centering
\includegraphics[width=8cm]{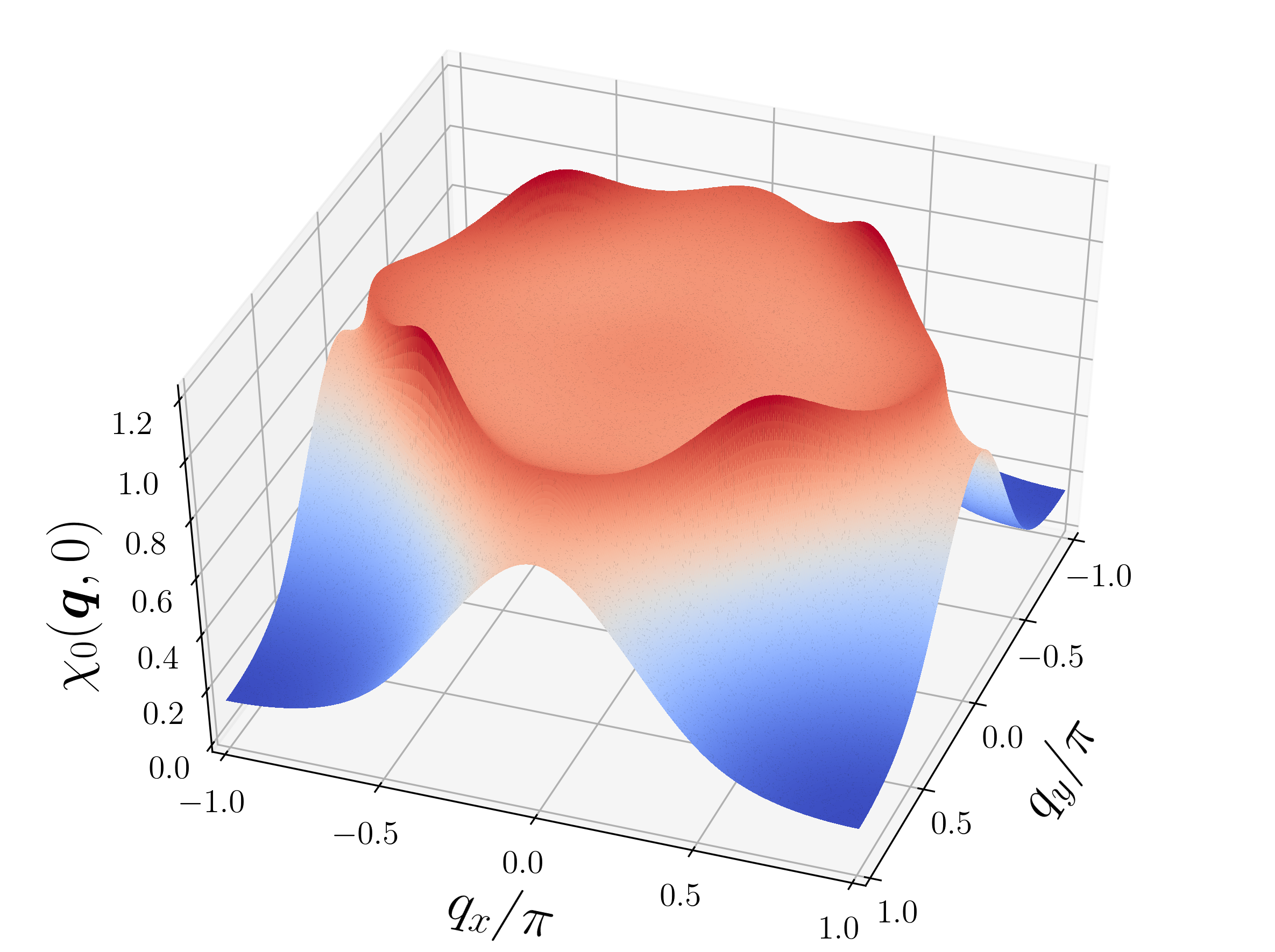}
\caption{Static bare susceptibility $\chi_0(\bq,0)$ as a function of $\bq$ for a tight-binding model of spin-$\frac{1}{2}$ fermions ($N=2$) with parameters as in Fig.~1.}
\end{figure}
In Fig.~2 we show the static (zero frequency) bare susceptibility $\chi_0(\bq,0) = -2 \Pi_0(\bq,0)$ for a tight-binding model with parameters as in Fig.~1 \cite{bonetti}. Pronounced peaks are visible at the nesting vectors connecting the four flat hot spots on the Fermi surface.

The RPA susceptibility diverges when $g\chi_0(\bQ,0) = -1$, signalling the onset of charge or spin density wave order with the nesting wave vector $\bQ$. To analyze the behavior of the susceptibility near the singularity, we expand
\begin{equation}
 \delta\Pi_0(\bq,\om) = \Pi_0(\bq,\om) - \Pi_0(\bQ,0) \, .
\end{equation}
for $\bq$ near $\bQ$ and small $\om$. Momenta near $\bQ$ are parametrized by relative momentum coordinates $q_r$ and $q_t$, parallel and perpendicular to $\bQ$, respectively.
The leading contributions to $\delta\Pi_0(\bq,\om)$ come from fermionic momenta near the hot spots connected by $\bQ$, where the dispersion relations in Eq.~\eqref{Pi0def} can be expanded as in Eq.~\eqref{dispersion}, that is, $\xi_\bk = v_F k_r + b k_t^4$ and
$\xi_{\bk-\bq} = - v_F (k_r - q_r) + b (k_t - q_t)^4$.
In the following we assume that $b$ is positive. Our derivations and results can be easily adapted to negative $b$.
The integrals over $k_r$, $k_t$, and $k_0$ are evaluated in Appendix~\ref{app:A}.

For $q_t=0$, the integral in Eq.~\eqref{Pi0def} is elementary, yielding
\begin{equation} \label{dPi0qt0}
 \delta\Pi_0(q_r,0,\om) =
 \frac{1}{4\pi v_F (2b)^{1/4}} \left[
 (1-i) \sqrt[4]{\om + i0^+ - v_F q_r} + 
 (1+i) \sqrt[4]{-\om - i0^+ - v_F q_r} \, \right] \, .
\end{equation}
In the static limit $\om \to 0$, one obtains
\begin{equation}
 \delta\Pi_0(q_r,0,0) = \left\{ \begin{array}{ll}
 (v_F |q_r|)^{1/4}/[2\pi v_F (2b)^{1/4}] & \mbox{for} \; q_r < 0 \, , \\
 (v_F q_r)^{1/4}/[\sqrt{2}\pi v_F (2b)^{1/4}] & \mbox{for} \; q_r > 0 \, .
 \end{array} \right.
\end{equation}
The static particle-hole bubble thus exhibits a cusp with infinite slope as a function of $q_r$ at $q_r = q_t = 0$.

\begin{figure}
\centering
\includegraphics[width=8cm]{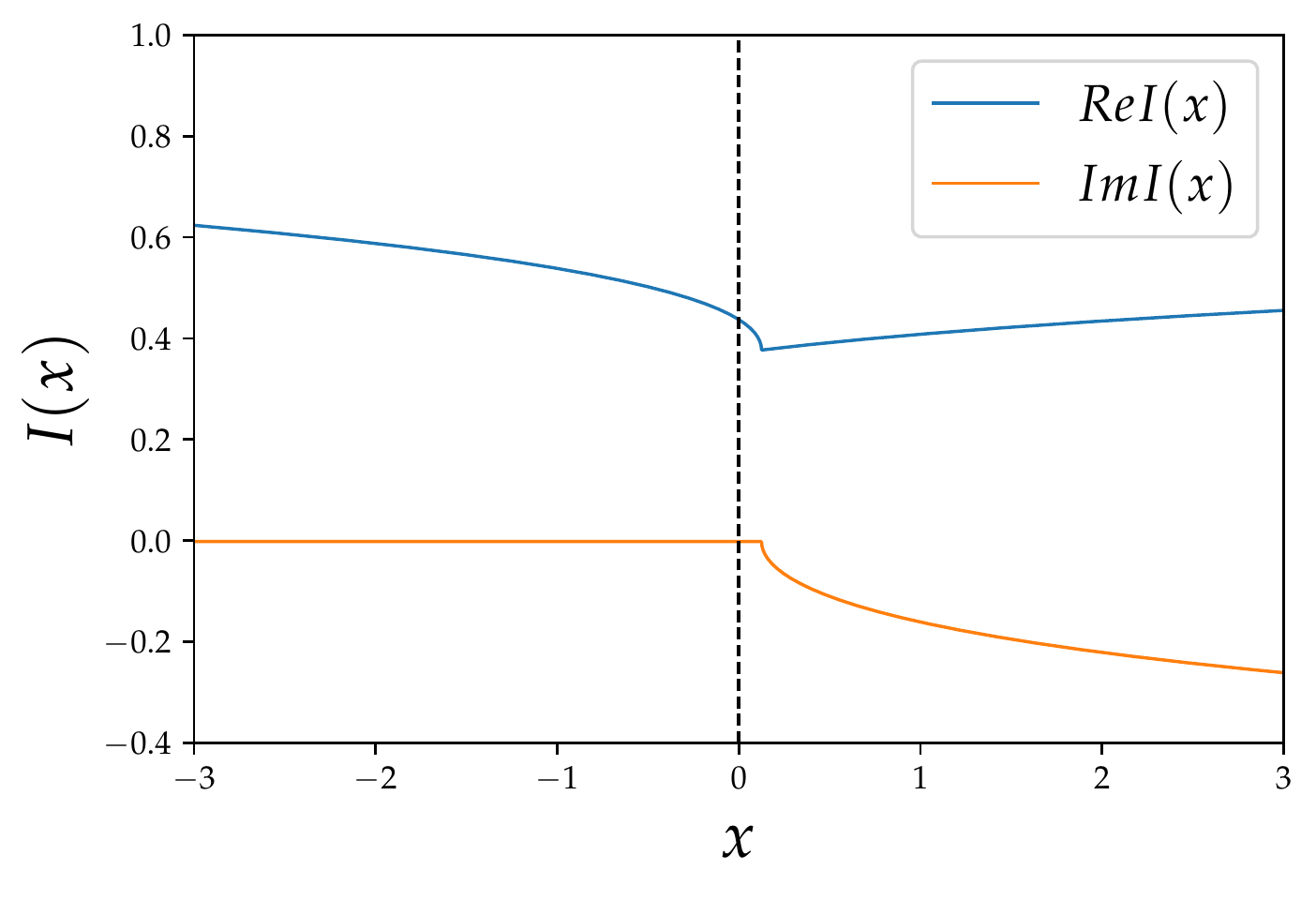}
\caption{Real and imaginary parts of the scaling function $I(x)$.}
\end{figure}
For $q_t \neq 0$, the particle-hole bubble can be expressed in a scaling form as (see Appendix~\ref{app:A})
\begin{equation} \label{dPi0}
 \delta\Pi_0(q_r,q_t,\om) = \frac{|q_t|}{4v_F} \, \Big[
 I\Big( \frac{\om - v_Fq_r}{bq_t^4} \Big) +
 I^*\Big( \frac{-\om - v_Fq_r}{bq_t^4} \Big) \Big] \, ,
\end{equation}
where
\begin{equation} \label{Ix}
 I(x) = \int_0^\infty \frac{d\tk_0}{2\pi} \left[
 \frac{1+i}{2 (i\tk_0)^{3/4}} -
 \frac{\frac{1}{\sqrt{-\frac{3}{4} + \sqrt{ \frac{1}{2} (1+x + 2i\tk_0)}}} +
 \frac{1}{\sqrt{-\frac{3}{4} - \sqrt{ \frac{1}{2} (1+x + 2i\tk_0)}}}}
 {\sqrt{2(1+x + 2i\tk_0)}} \right]
\end{equation}
is a dimensionless scaling function -- shown graphically in Fig.~3. The graph of $I(x)$ exhibits a cusp at $x=\frac{1}{8}$, and the imaginary part of $I(x)$ vanishes for all $x \leq \frac{1}{8}$. While $I(x)$ cannot be expressed by elementary functions, $I(0)$ and $I\left(\frac{1}{8}\right)$ are given by the simple numbers $\sqrt{2}/\pi$ and $\sqrt{\frac{3}{2}}/\pi$, respectively. Since $I(0)$ is finite, $\delta\Pi_0(0,q_t,0)$ is linear in $|q_t|$.
For $|x| \to \infty$, the scaling function behaves asymptotically as
\begin{equation}
 I(x) \sim \left\{ \begin{array}{lll}
 \frac{1-i}{2^{1/4}\pi} \, x^{1/4} &+ \; \frac{3 \cdot 2^{1/4}(1+i)}{8\pi} \, x^{-1/4} 
 & \mbox{for} \; x > 0 \, , \\
 \frac{2^{1/4}}{\pi} \, |x|^{1/4} &+ \; \frac{3}{4 \cdot 2^{1/4} \pi} \, |x|^{-1/4} 
 & \mbox{for} \; x < 0 \, .
 \end{array} \right.
\end{equation}
Inserting the leading asymptotic behavior into Eq.~\eqref{dPi0} one recovers the result Eq.~\eqref{dPi0qt0} for $q_t=0$. The next to leading order yields the leading $q_t$ dependence for $b|q_t|^4 \ll |\pm\om - v_Fq_r|$,
\begin{equation}
 \Pi_0(q_r,q_t,\om) - \Pi_0(q_r,0,\om) =
 \frac{3(2b)^{1/4} q_t^2}{32\pi v_F} \left(
 \frac{1+i}{\sqrt[4]{\om + i0^+ - v_Fq_r}} +
 \frac{1-i}{\sqrt[4]{-\om - i0^+ - v_Fq_r}} \right) + \cO(q_t^4) \, .
\end{equation}
Hence, for $\om \neq \pm v_Fq_r$, the leading $q_t$ dependence is quadratic in $q_t$.

The RPA effective interaction is given by
\begin{equation} \label{D_rpa}
 D(\bq,iq_0) = \frac{g}{1 + g \chi_0(\bq,iq_0)}
\end{equation}
on the imaginary frequency axis, and by the same expression with $iq_0 \to \om$ on the real frequency axis. At the QCP, $g \chi_0(\bQ,0)$ is equal to minus one, so that
\begin{equation} \label{D_qcp}
 D(\bq,\om) = - \frac{1}{N \delta\Pi_0(\bq,\om)} \, .
\end{equation}
Hence, the effective interaction at the QCP does not depend on the coupling constant $g$.


\section{Fermion self-energy}

To leading order in the effective interaction $D$, the fermion self-energy is given by the one-loop integral
\begin{equation}
 \Sg(\bk,ik_0) = - M \int_\bq \int_{q_0} D(\bq,iq_0) \,
 G_0(\bk-\bq,ik_0-iq_0) \, ,
\end{equation}
with $M=1$ for a charge density and $M=3$ for a spin-density instability \cite{sykora18}. 
Analytic continuation of this expression to the real frequency axis yields \cite{rohe01}
\begin{eqnarray} \label{Sigma_realfreq}
 \Sg(\bk,\om+i0^+) &=& - \frac{M}{\pi} \int d\nu \int_\bq \big[
 b(\nu) \Im D(\bq,\nu+i0^+) \, G_0(\bk-\bq,\nu+\om+i0^+) \nonumber \\
 && - \, f(\nu) \, D(\bq,\nu-\om-i0^+) \, \Im G_0(\bk-\bq,\nu+i0^+) 
 \big] \, ,
\end{eqnarray}
where $b(\nu) = [e^{\beta\nu} - 1]^{-1}$ and $f(\nu) = [e^{\beta\nu} + 1]^{-1}$ are the Bose and Fermi functions, respectively. At zero temperature ($\beta = \infty$) these functions become step functions $b(\nu) = - \Theta(-\nu)$ and $f(\nu) = \Theta(-\nu)$.
In the following we denote $\Sg(\bk,\om+i0^+)$, $G(\bk,\om+i0^+)$, and $D(\bq,\nu+i0^+)$ by $\Sg(\bk,\om)$, $G(\bk,\om)$, and $D(\bq,\nu)$, respectively.
Note, however, the negative infinitesimal imaginary part in one of the frequency arguments in Eq.~\eqref{Sigma_realfreq}.

We analyze $\Sg(\bk,\om)$ at the QCP for low frequencies $\om$ and momenta $\bk$ near one of the hot spots on the Fermi surface, which we denote as $\bk_H$. The effective interaction $D$ at the QCP is given by Eq.~\eqref{D_qcp} with $\delta\Pi_0$ from Eq.~\eqref{dPi0}.
The dominant contributions come from momentum transfers $\bq$ near $\bQ$, so that $\bk-\bq$ is situated near the antipodal hot spot $-\bk_H$. Using relative momentum variables as above, the dispersion relation in the fermion propagator can be expanded as $\xi_{\bk-\bq} = - v_F (k_r-q_r) + b (k_t-q_t)^4$.

To evaluate the self-energy, it is convenient to first consider its imaginary part, and then compute the real part from a Kramers-Kronig relation. The imaginary part of Eq.~\eqref{Sigma_realfreq} reads
\begin{equation} \label{ImSigma1}
 \Im\Sg(\bk,\om) = - \frac{M}{\pi} \int d\nu \int_\bq
 \left[ b(\nu) + f(\nu+\om) \right] \,
 \Im D(\bq,\nu) \, \Im G_0(\bk-\bq,\om+\nu) \, .
\end{equation}
Note that $\Im D(\bq,\nu-i0^+) = - \Im D(\bq,\nu+i0^+)$.
Using the Dirac identity $\Im G_0(\bk,\om) = -\pi \delta(\om-\xi_{\bk})$, the frequency integral in Eq.~\eqref{ImSigma1} can be easily carried out, yielding
\begin{equation} \label{ImSigma2}
 \Im\Sg(\bk,\om) = M \int_\bq
 \left[ b(\xi_{\bk-\bq} - \om) + f(\xi_{\bk-\bq}) \right] \,
 \Im D(\bq,\xi_{\bk-\bq} - \om) \, .
\end{equation}
At zero temperature, the sum of Bose and Fermi functions in Eq.~\eqref{ImSigma2} is given by
\begin{equation}
 b(\xi_{\bk-\bq} - \om) + f(\xi_{\bk-\bq}) = \left\{ \begin{array}{rl}
 -1 & \mbox{for} \; 0 < \xi_{\bk-\bq} < \om \, , \\
  1 & \mbox{for} \; \om < \xi_{\bk-\bq} < 0 \, , \\
  0 & \mbox{else} \, , \end{array} \right.
\end{equation}
restricting thus the contributing momentum region.
The integral in Eq.~\eqref{ImSigma2} is convergent even if the momentum integration over $q_r$ and $q_t$ is extended to infinity.

The real part of the self-energy can be obtained from the Kramers-Kronig-type relation
\begin{equation} \label{KK}
 \Sg(\bk,\om) = - \frac{1}{\pi} \int_{-\infty}^{\infty} d\om' \,
 \frac{\Im\Sg(\bk,\om')}{\om - \om' + i0^+} + \mbox{const} \, .
\end{equation}
The last term in this relation is a real constant.


\subsection{Frequency dependence at hot spot}

The frequency dependence at the hot spot (for $\bk=\bk_H$) can be derived by a simple rescaling of the integration variables in Eq.~\eqref{ImSigma2}. Substituting $q_r = |\om/v_F| \tq_r$ and $q_t = |\om/b|^{1/4} \tq_t$, one obtains
\begin{equation} \label{ImSigma3}
 \Im\Sg(\bk_H,\om) = - \frac{M}{N} \, A_{s(\om)} |\om| \, ,
\end{equation}
where $A_+$ and $A_-$ are two positive dimensionless numbers depending on the sign of $\om$. These numbers are determined by the integral
\begin{equation} \label{Apm}
 A_{s(\om)} = - \int'_{\tilde\bq} \Im \frac{4 s(\om)}
 {|\tq_t| \left[ I \left( \frac{\tq_t^4 - s(\om)}{\tq_t^4} \right) +
 I^* \left( \frac{-2\tq_r - \tq_t^4 + s(\om)}{\tq_t^4} \right) \right]} \, ,
\end{equation}
where the prime at the integral sign indicates a restriction of the integration region to $0 < \tq_r + \tq_t^4 < 1$ for $\om > 0$, and to $-1 < \tq_r + \tq_t^4 < 0$ for $\om < 0$.
Note that the frequency dependence of the self-energy at the hot spot depends neither on $v_F$, nor on $b$.
A numerical evaluation of the integral in Eq.~\eqref{Apm} yields $A_+ \approx 0.049$ and $A_- \approx 0.072$. Hence, $\Im\Sg(\bk_H,\om)$ is slightly asymmetric in $\omega$.
For the generic situation with a finite Fermi surface curvature, a linear frequency dependence of $\Im\Sg(\bk_H,\om)$ with asymmetric coefficients has also been found in a one-loop calculation of the self-energy at the onset of charge or spin-density wave order with {\em two}\/ pairs of nested hot spots connected by the same wave vector $\bQ$ \cite{holder14,sykora21}. In that case, however, the coefficients are not universal -- they depend on a dimensionless combination of model parameters.

The real part of the self-energy can be obtained from the Kramers-Kronig relation Eq.~\eqref{KK}. With $\Im\Sg(\bk_H,\om)$ as in Eq.~\eqref{ImSigma3}, the integral in Eq.~\eqref{KK} is logarithmically divergent at large frequencies $\om'$. This is due to the fact that the linear frequency dependence has been obtained from an expansion that captures only the asymptotic low frequency behavior, which cannot be extended to all frequencies. The imaginary part of the exact self-energy of any physical system has to vanish in the high-frequency limit. To compute the low-frequency behavior of $\Re\Sg$, we mimic the high-frequency decay of $\Im\Sg$ by imposing an ultraviolet frequency cutoff $\Lam_\om$, so that the frequency integration in Eq.~\eqref{KK} is restricted to $|\om'| < \Lam_\om$. Defining $\delta\Sg(\bk,\om) = \Sg(\bk,\om) - \Sg(\bk_H,0)$, we then obtain
\begin{equation} \label{ReSg_om}
 \Re \, \delta\Sg(\bk_H,\om) =
 - \frac{M}{N} \, \frac{A_+ + A_-}{\pi} \, \om \ln\frac{\Lam_\om}{|\om|} \, ,
\end{equation}
for $|\om| \ll \Lam_\om$.

The logarithm in Eq.~\eqref{ReSg_om} implies a logarithmic divergence of the inverse quasiparticle weight \cite{negele87},
$Z^{-1} = 1 - \partial\Sg(\bk_H,\om)/\partial\om \sim \ln(\Lam_\om/|\om|)$.
Hence, Landau quasiparticles do not exist at the hot spots, and Fermi liquid theory breaks down.

Logarithmic divergences are frequently a perturbative manifestation of power-law behavior, especially in (quantum) critical systems. Assuming that the one-loop result in Eq.~\eqref{ReSg_om} reflects the leading order of an expansion of a power-law, one obtains
\begin{equation} \label{Sg_power}
 \om - \delta\Sg(\bk_H,\om) \propto \big( |\om|/\Lam_\om \big)^{-\eta_\om} \om
\end{equation}
at low frequencies, with the anomalous dimension
\begin{equation}
 \eta_\om = \frac{M}{N} \, \frac{A_+ + A_-}{\pi} \approx \frac{M}{N} \, 0.039 \, .
\end{equation}
Hence, the quasiparticle weight $Z$ vanishes as $|\om|^{\eta_\om}$ in the low-energy limit.
Note that Kramers-Kronig consistency requires that the real and imaginary parts of the self-energy obey the same power-law if $\eta_\om > 0$.
We emphasize that the power-law in Eq.~\eqref{Sg_power} is only an educated guess. The actual behavior might be more complicated. To clarify the role of higher order contributions, a renormalization group analysis might be a useful next step.


\subsection{Frequency and momentum dependencies near hot spot}

We now analyze the momentum and frequency dependence of the self-energy in the vicinity of a hot spot. We consider radial and tangential momentum dependencies separately.

\begin{figure}
\centering
\includegraphics[width=8cm]{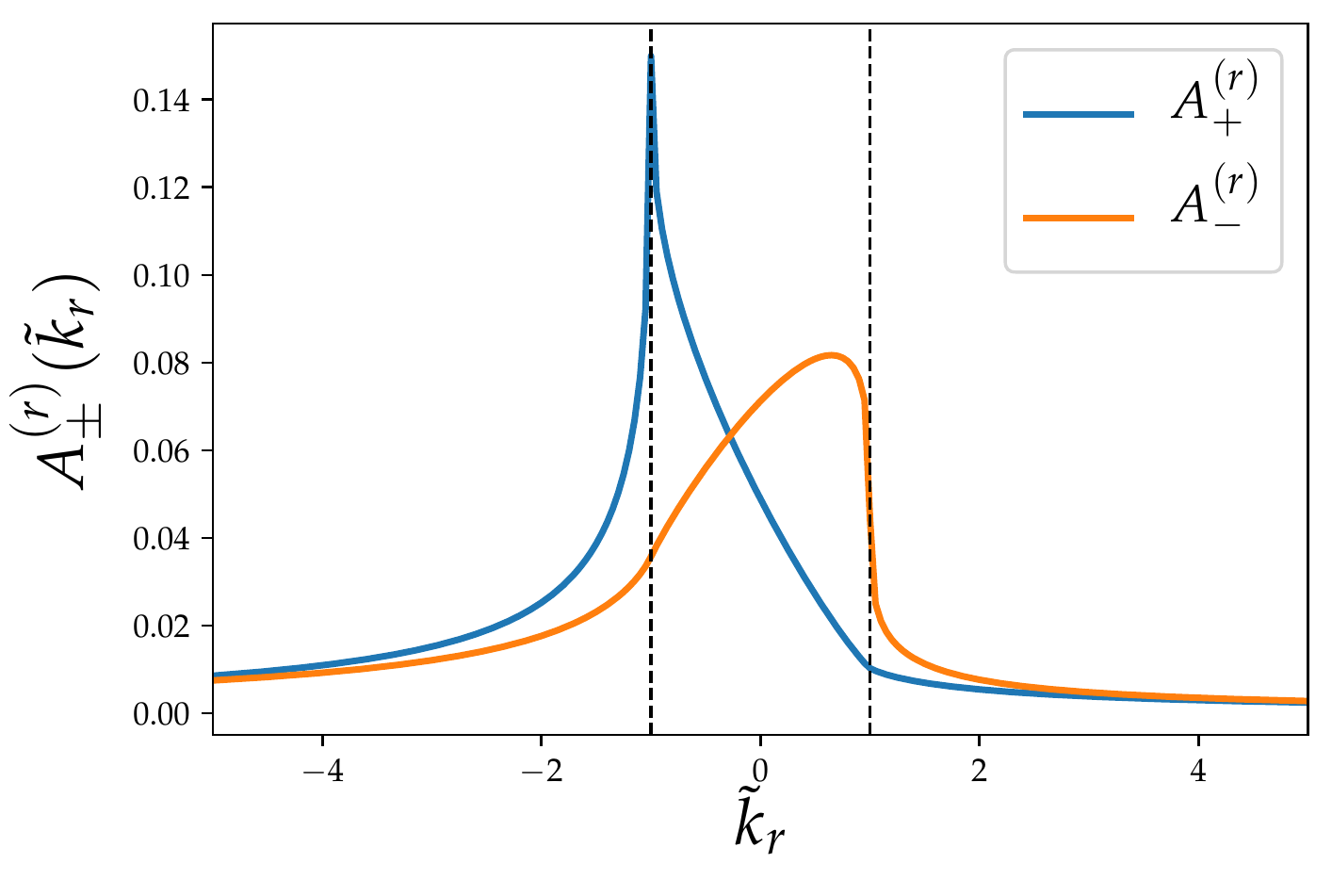}
\caption{Scaling functions $A^{(r)}_\pm(\tk_r)$. The vertical dashed lines mark the location of singularities at $\tk_r = \pm 1$.}
\end{figure}
For $k_t = 0$, we can express $\Im\Sg(\bk,\om)$ from Eq.~\eqref{ImSigma2} in the scaling form
\begin{equation} \label{ImSigma4}
 \Im\Sg(\bk,\om) = - \frac{M}{N} \, A^{(r)}_{s(\om)}(\tk_r) \, |\om| \, , 
\end{equation}
with the dimensionless scaling functions
\begin{equation} \label{Apmr}
 A^{(r)}_{s(\om)}(\tk_r) = - \int'_{\tilde\bq} \Im \frac{4 s(\om)}
 {|\tq_t| \left[ I \left( \frac{-\tk_r + \tq_t^4 - s(\om)}{\tq_t^4} \right) +
 I^* \left( \frac{\tk_r - 2\tq_r - \tq_t^4 + s(\om)}{\tq_t^4} \right) \right]} \, ,
\end{equation}
where the integration region is restricted to $0 < - \tk_r + \tq_r + \tq_t^4 < 1$ for $\om > 0$, and to $-1 < - \tk_r + \tq_r + \tq_t^4 < 0$ for $\om < 0$.
The rescaled variables are defined by $q_r = |\om/v_F| \tq_r$, $q_t = |\om/b|^{1/4} \tq_t$, and $k_r = |\om/v_F| \tk_r$.
The scaling functions $A_\pm^{(r)}$ are shown graphically in Fig.~4.
For $\tk_r = 0$ we recover Eq.~\eqref{ImSigma3}, since $A^{(r)}_\pm(0) = A_\pm$ from Eq.~\eqref{Apm}. For small finite $\tk_r$, the leading $\tk_r$ dependence of $A^{(r)}_{s(\om)}(\tk_r)$ is linear,
\begin{equation} \label{Apmr_asymp}
 A^{(r)}_\pm(\tk_r) = A_\pm + B_\pm \tk_r + \cO(\tk_r^2) \, ,
\end{equation}
with $B_+ \approx - 0.050$ and $B_- \approx 0.027$.

\begin{figure}
\centering
\includegraphics[width=8cm]{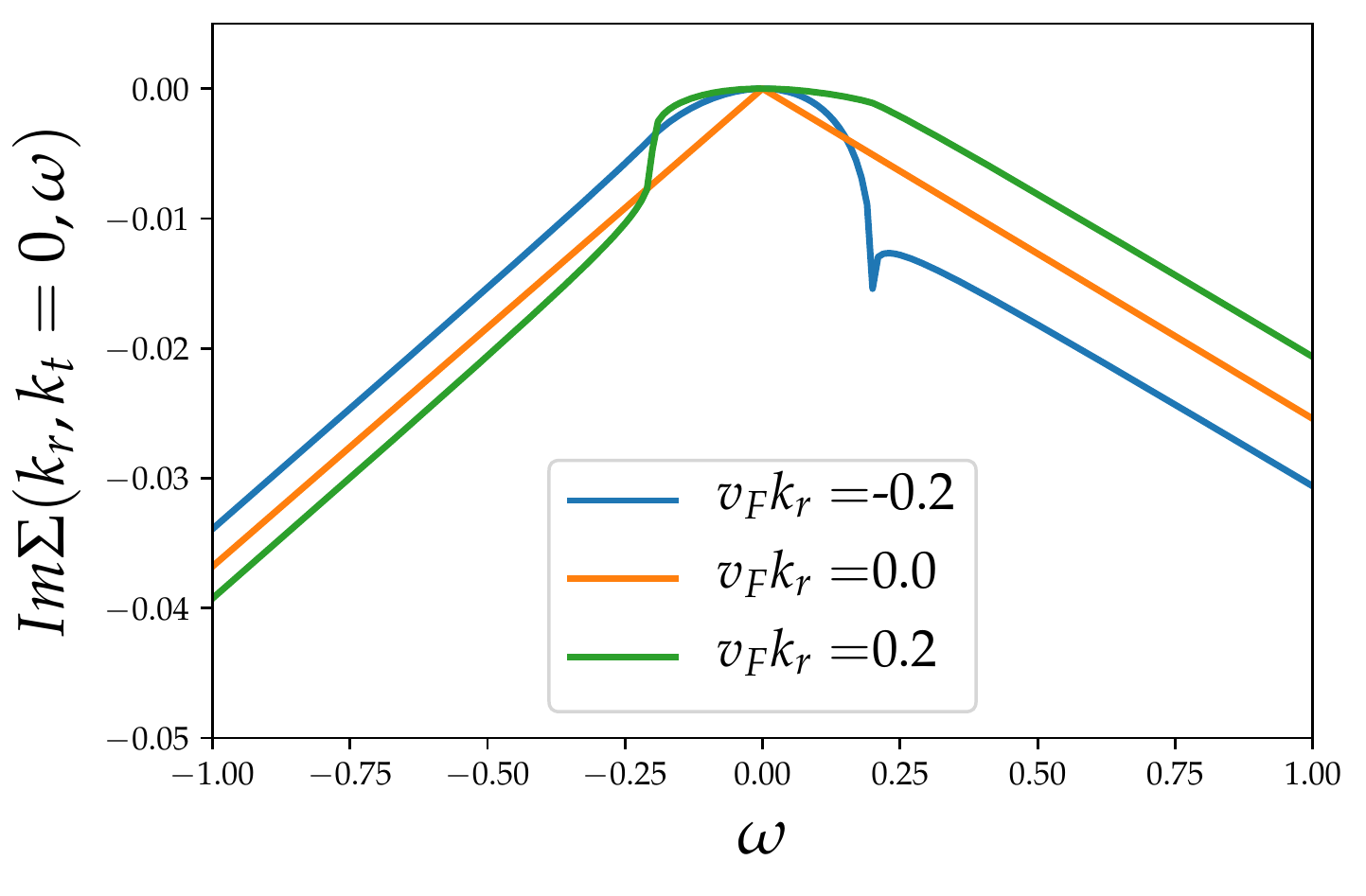}
\caption{Imaginary part of the one-loop self-energy as a function of frequency for various choices of the radial momentum variable $k_r$. Here $M=1$ and $N=2$.}
\end{figure}
In Fig.~5 we show the frequency dependence of $\Im\Sg(\bk,\om)$ for various choices of $k_r$. For small $|\om|$, the leading frequency dependence is quadratic. For $|\om| \gg v_F |k_r|$, the curves rapidly approach the asymptotic behavior
\begin{equation} \label{ImSigma5}
 \Im\Sg(\bk,\om) \sim - \frac{M}{N} \left[ A_{s(\om)} |\om| + B_{s(\om)} v_F k_r \right] \, ,
\end{equation}
which follows from Eq.~\eqref{Apmr_asymp}.
Inserting this asymptotic dependence in the Kramers-Kronig relation Eq.~\eqref{KK}, one obtains the leading $k_r$ dependence of the real part of the self-energy at zero frequency as
\begin{equation}
 \delta\Sg(\bk,0) \sim - \, \frac{M}{N} \, \frac{B_+ - B_-}{\pi} \, v_F k_r \,
 \ln\frac{\Lam_\om}{v_F |k_r|}  \, .
\end{equation}
Assuming, as before, that the logarithm reflects the leading contribution from a power-law, we expect a momentum dependence of the form
\begin{equation} \label{Sg_power2}
 v_F k_r + \delta\Sg(\bk,0) \propto \big( v_F|k_r|/\Lam_\om \big)^{-\eta_r} v_F k_r
\end{equation}
for small $k_r$, with the anomalous dimension
\begin{equation}
 \eta_r = \frac{M}{N} \, \frac{B_- - B_+}{\pi} \approx \frac{M}{N} \, 0.025  \, .
\end{equation}
The renormalized Fermi velocity \cite{negele87} given by
$\bar v_F(k_r) = Z \left[ 1 + \partial\Sg/\partial k_r \right] v_F$ is proportional to
$|\om|^{\eta_\om} |k_r|^{-\eta_r}$ with $\om = v_F k_r$, and thus
$\bar v_F(k_r) \propto |k_r|^{\eta_\om - \eta_r}$.
This quantity vanishes for $k_r \to 0$, albeit very slowly, since $\eta_\om > \eta_r$.

\begin{figure}
\centering
\includegraphics[width=8cm]{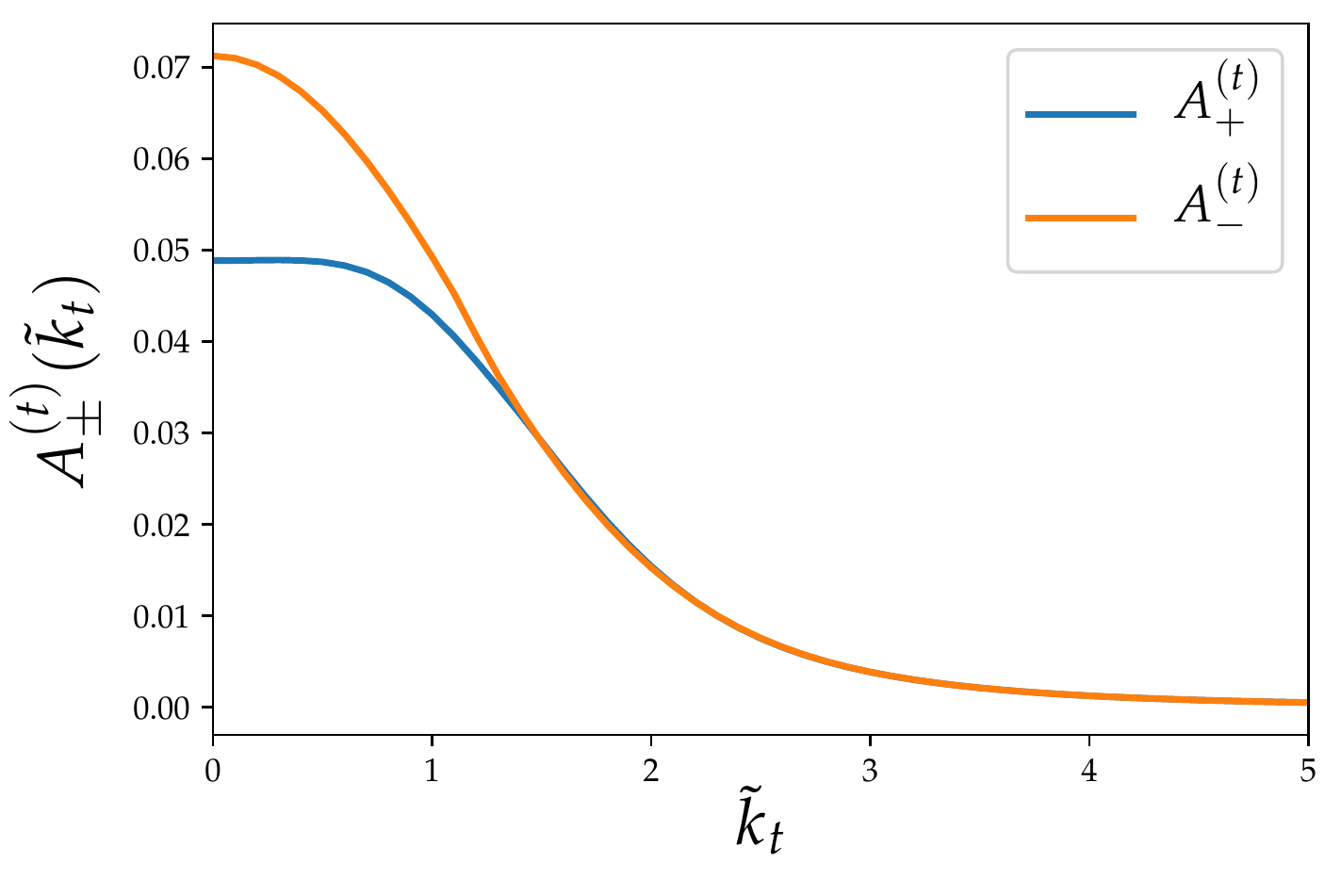}
\caption{Scaling functions $A^{(t)}_\pm(\tk_t)$.}
\end{figure}
We now discuss the tangential momentum dependence of the self-energy.
For $k_r = 0$, we can express $\Im\Sg(\bk,\om)$ from Eq.~\eqref{ImSigma2} in the scaling form
\begin{equation} \label{ImSigma6}
 \Im\Sg(\bk,\om) = - \frac{M}{N} \, A^{(t)}_{s(\om)}(\tk_t) \, |\om| \, , 
\end{equation}
with the dimensionless scaling functions (see also Fig.~6)
\begin{equation} \label{Apmt}
 A^{(t)}_{s(\om)}(\tk_t) = - \int'_{\tilde\bq} \Im \frac{4 s(\om)}
 {|\tq_t| \left[ I \left( \frac{(\tk_t-\tq_t)^4 - s(\om)}{\tq_t^4} \right) +
 I^* \left( \frac{- 2\tq_r - (\tk_t-\tq_t)^4 + s(\om)}{\tq_t^4} \right) \right]} \, ,
\end{equation}
where the integration region is restricted to $0 < \tq_r + (\tk_t-\tq_t)^4 < 1$ for $\om > 0$, and to $-1 < \tq_r + (\tk_t-\tq_t)^4 < 0$ for $\om < 0$.
The rescaled variables are defined by $q_r = |\om/v_F| \tq_r$, $q_t = |\om/b|^{1/4} \tq_t$, and $k_t = |\om/b|^{1/4} \tk_t$.
The scaling functions $A^{(t)}_\pm(\tk_t)$ are symmetric under $\tk_t \mapsto -\tk_t$. Hence, their Taylor expansion for small $\tk_t$ contains only even powers of $\tk_t$,
\begin{equation}
 A^{(t)}_\pm(\tk_t) = A_\pm + C_\pm \tk_t^2 + D_\pm \tk_t^4 + \cO(\tk_t^6) \, .
\end{equation}
A numerical evaluation of Eq.~\eqref{Apmt} yields $C_+ = 0$, $C_- \approx - 0.023$, and $D_+ \approx - 0.008$. It is difficult to extract the quartic coefficient $D_-$ numerically, because the quartic contribution is superseded by the much larger (for small $\tk_t$) quadratic term.

\begin{figure}
\centering
\includegraphics[width=8cm]{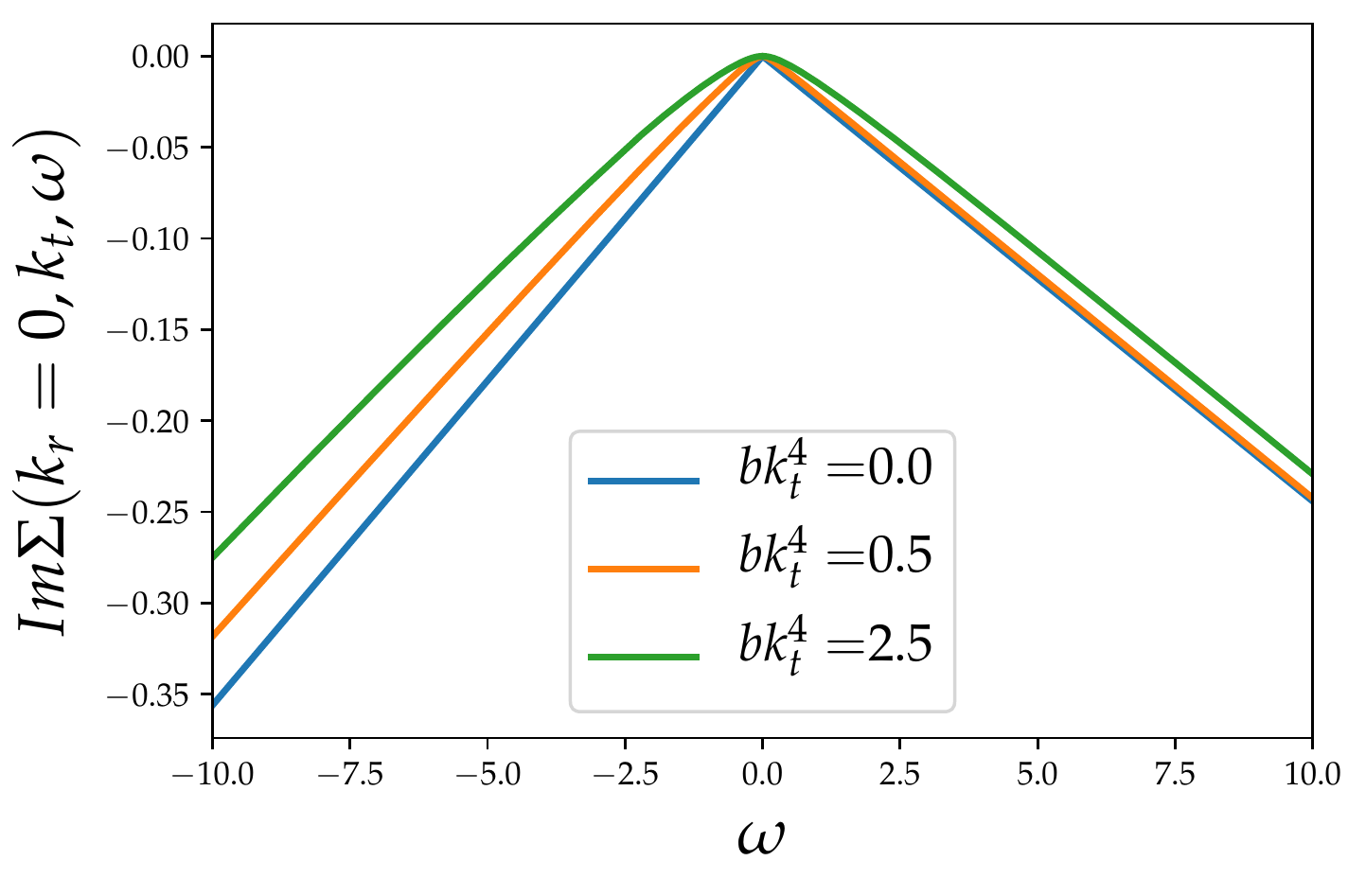}
\caption{Imaginary part of the one-loop self-energy as a function of frequency for various choices of the tangential momentum variable $k_t$. Here $M=1$ and $N=2$.}
\end{figure}
In Fig.~7 we show the frequency dependence of $\Im\Sg(\bk,\om)$ for various choices of $k_t$. For small $|\om|$, the leading frequency dependence is quadratic. For $|\om| \gg b |k_t|^4$, the curves approach the asymptotic behavior
\begin{equation}
 \Im\Sg(\bk,\om) \sim - \frac{M}{N} \left[ 
 A_{s(\om)} |\om| + C_{s(\om)} \sqrt{b|\om|} k_t^2 + D_{s(\om)} bk_t^4 \right] \, .
\end{equation}
Inserting this expansion into the Kramers-Kronig relation Eq.~\eqref{KK}, we obtain
\begin{equation} \label{ReSigma_kt}
 \delta\Sg(\bk,0) = - \frac{2}{\pi} (C_+ - C_-) \sqrt{b\Lam_\om} \, k_t^2 -
 \frac{D_+ - D_-}{\pi} \, bk_t^4 \ln\frac{\Lam_\om}{bk_t^4} + \cO(k_t^4) \, ,
\end{equation}
for small $k_t$. The leading term is quadratic in $k_t$ and depends strongly on the ultraviolet cutoff $\Lam_\om$. This term, along with generic regular many-body contributions of the same order, leads to a renormalized dispersion relation $\bar\xi_\bk$ with a quadratic tangential momentum dependence, in conflict with our original assumption. However, the case of a dispersion with a vanishing quadratic dependence on $k_t$ can be restored by slightly shifting the bare parameters of our system such that the self-energy corrections cancel the quadratic $k_t$ dependence of the bare dispersion relation. Between parameter regimes with a locally convex and a locally concave Fermi surface (at the hot spots), there must be a transition point where the Fermi surface is flat, in the interacting system as well as in the non-interacting reference model.
Hence, we are left with the second term in Eq.~\eqref{ReSigma_kt} only, which can be promoted to a power-law of the form
\begin{equation} \label{Sg_power3}
 bk_t^4 + \delta\Sg(\bk,0) \propto \big( bk_t^4/\Lam_\om \big)^{-\eta_t} bk_t^4
\end{equation}
for small $k_t$, with the anomalous dimension
\begin{equation}
 \eta_t = \frac{M}{N} \, \frac{D_- - D_+}{\pi} \, .
\end{equation}
We have no estimate of the number for $D_-$, but from the numerical data for $A_-(\tk_t)$ we can see that it must be very small, such that the anomalous dimension $\eta_t$ is also small, in line with the other anomalous dimensions $\eta_\om$ and $\eta_r$.

Combining the results on the frequency dependence of the quasiparticle weight $Z$ with the results on the radial and tangential momentum dependence of the self-energy, we obtain a renormalized dispersion relation of the form
\begin{equation} \label{xi_ren}
 \bar\xi_\bk = a_r \, {\rm sgn}(k_r) |k_r|^{\alpha_r} + a_t |k_t|^{\alpha_t} \, ,
\end{equation}
where $\alpha_r = 1 + \eta_\om - \eta_r$ and $\alpha_t = 4(1 + \eta_\om - \eta_t)$, while $a_r$ and $a_t$ are non-universal coefficients. The anomalous dimensions are relatively small, so that $\alpha_r$ and $\alpha_t$ remain close to the bare values one and four, respectively. A renormalized dispersion of the form Eq.~\eqref{xi_ren} has been obtained earlier in an $\epsilon$-expansion \cite{halbinger19} for the standard case of a bare dispersion with a quadratic tangential momentum dependence. In that case, the anomalous dimensions turned out to be quite large in the physical dimension two (corresponding to $\epsilon = \frac{1}{2}$), leading to a strongly flattened Fermi surface with an almost quartic shape $k_r \propto |k_t|^{3.85}$.


\section{Conclusion}

We have analyzed quantum fluctuation effects at the onset of charge or spin density wave order with an incommensurate nesting ($2k_F$) wave vector $\bQ$ in two-dimensional metals -- for the special case where $\bQ$ connects a pair of hot spots situated at flat high symmetry points of the Fermi surface with a vanishing Fermi surface curvature. The leading tangential momentum dependence of the bare dispersion is quartic at these points. The charge or spin susceptibilities form a pronounced peak at $\bQ$.

We have computed the fermion self-energy $\Sg(\bk,\om)$ at the QCP as a function of (real) frequency and momentum near the hot spots in RPA, that is, in one-loop approximation. At the hot spots, the frequency dependence of $\Im\Sg$ is linear and slightly asymmetric, while the quasiparticle weight and the momentum dependence of the self-energy exhibit logarithmic divergences with universal prefactors. Hence, there are no Landau quasiparticles at the hot spots, giving rise to non-Fermi liquid behavior.

A tentative resummation of the logarithms leads to power-laws with small universal anomalous dimensions. The quasiparticle weight vanishes with a small power of frequency. The renormalized dispersion relation has the form
$\bar\xi_\bk = a_r \, {\rm sgn}(k_r) |k_r|^{\alpha_r} + a_t |k_t|^{\alpha_t}$
near the hot spots, where $k_r$ and $k_t$ are radial and tangential relative momentum variables, respectively. The exponents $\alpha_r$ and $\alpha_t$ are close to the corresponding exponents of the bare dispersion relation one and four, respectively.
Since the renormalized dispersion relation has almost the same form as the bare one, and the quasiparticle weight vanishes only slowly, the self-energy corrections do not destroy the peak at the nesting vector in the susceptibility. The $2k_F$ QCP is thus stable. In particular, there are no indications that it might be replaced by a first order transition, in contrast to the more delicate situation for a quadratic dispersion \cite{altshuler95}.

The QCP with flat hot spots could be realized experimentally in suitable layered compounds or cold atom systems by fine tuning two parameters, for example, density and interaction strength.
Moreover, our model is an instructive prototype of a larger class of systems with a {\em bare} dispersion of the form $\xi_\bk = a_r \, {\rm sgn}(k_r) |k_r|^{\alpha_r^{(0)}} + a_t |k_t|^{\alpha_t^{(0)}}$. In our case, where $\alpha_r^{(0)}= 1$ and $\alpha_t^{(0)} = 4$, the analysis is comparatively simple, since several integrations can be performed analytically, and the renormalized dispersion remains close to the bare one.
From a theoretical point of view it would be interesting to extend the analysis to more general bare exponents $\alpha_r^{(0)}$ and $\alpha_t^{(0)}$, and to see whether the renormalized dispersion has universal exponents. In particular, it remains to be clarified whether in the conventional but delicate case $\alpha_r^{(0)} = 1$ and $\alpha_t^{(0)} = 2$ the mean-field QCP survives in the presence of fluctuations, with a renormalized dispersion of the above form and a flattened Fermi surface, as suggested by the one-loop $\epsilon$-expansion by Halbinger et al.\ \cite{halbinger19}.


\begin{acknowledgments}
We are grateful to Walter Hofstetter, Thomas Sch\"afer, and J\'achym S\'ykora for valuable discussions, and to Pietro Bonetti for providing the data for the bare susceptibility shown in Fig.~2.
\end{acknowledgments}


\begin{appendix}

\section{Evaluation of particle-hole bubble} \label{app:A}

The $k_r$ integral in Eq.~\eqref{Pi0def} can be easily carried out by using the residue theorem. Shifting the remaining integration variables as $k_t \to k_t + q_t/2$ and $k_0 \to k_0 + q_0/2$ to symmetrize the integrand, one obtains
\begin{equation} \label{Pi0krint}
 \Pi_0(\bq,iq_0) = \frac{i}{v_F}
 \int_{-\infty}^{\infty} \frac{dk_0}{2\pi} \int_{-\infty}^{\infty} \frac{dk_t}{2\pi} \,
 \frac{\Theta(-k_0-q_0/2) - \Theta(k_0-q_0/2)}
 {2ik_0 - v_F q_r - b(k_t + q_t/2)^4 - b(k_t - q_t/2)^4}
\end{equation}
The $k_t$ integration can also be performed via residues. The denominator in Eq.~\eqref{Pi0krint} has four poles in the complex $k_t$ plane, namely
\begin{equation}
 k_t^{ss'} = s \sqrt{ - \frac{3}{4} q_t^2 + \frac{s'}{\sqrt{2}}
 \sqrt{q_t^4 + (2ik_0 - v_F q_r)/b} } \, ,
\end{equation}
where $s,s' \in \{+,-\}$.
Closing the integration contour in the upper complex half-plane, only the poles in the upper half-plane contribute. For $k_0 > 0$, these are the poles $k_t^{++}$ and $k_t^{--}$, and the corresponding residues are
\begin{subequations} \label{residues}
\begin{align}
 R^{++} &= \frac{1}{(k_t^{++} - k_t^{+-})(k_t^{++} - k_t^{-+})(k_t^{++} - k_t^{--})}
 = \frac{1}{2\sqrt{2} k_t^{++} \sqrt{q_t^4 + (2ik_0 - v_F q_r)/b}} \, , \\
 R^{--} &= \frac{1}{(k_t^{--} - k_t^{++})(k_t^{--} - k_t^{+-})(k_t^{--} - k_t^{-+})}
 =  \frac{-1}{2\sqrt{2} k_t^{--} \sqrt{q_t^4 + (2ik_0 - v_F q_r)/b}} \, .
\end{align}
\end{subequations}
For $k_0 < 0$, the poles $k_t^{+-}$ and $k_t^{-+}$ are situated in the upper half-plane, and the corresponding residues are
\begin{subequations}
\begin{align}
 R^{+-} &= \frac{1}{(k_t^{+-} - k_t^{++})(k_t^{+-} - k_t^{-+})(k_t^{+-} - k_t^{--})}
 = - R^{--} \, , \\
 R^{-+} &= \frac{1}{(k_t^{-+} - k_t^{++})(k_t^{-+} - k_t^{+-})(k_t^{-+} - k_t^{--})}
 = - R^{++} \, ,
\end{align}
\end{subequations}
where $R^{++}$ and $R^{--}$ are defined by the expressions on the right hand sides of Eq.~\eqref{residues}, but now for $k_0 < 0$.
The numerator in Eq.~\eqref{Pi0krint} partitions the $k_0$ axis in three regions,
\begin{equation}
 \Theta(-k_0-q_0/2) - \Theta(k_0-q_0/2) = \left\{ \begin{array}{rll}
 1 & \mbox{for} & k_0 < - |q_0|/2 \\
 0 & \mbox{for} & - |q_0|/2 < k_0 < |q_0|/2 \\
 -1 & \mbox{for} & k_0 > |q_0|/2
 \end{array} \right.
\end{equation}
Performing the $k_t$ integral in Eq.~\eqref{Pi0krint} by using the residue theorem, one thus obtains
\begin{equation} \label{Pi0krktint}
 \Pi_0(\bq,iq_0) = - \frac{1}{2bv_F} \int_{-\infty}^{\infty} \frac{dk_0}{2\pi} \,
 \Theta(|k_0| - |q_0|/2) \left( R^{++} + R^{--} \right) \, .
\end{equation}
The integral diverges in the ultraviolet, but the integral for
$\delta\Pi_0(\bq,iq_0) = \Pi_0(\bq,iq_0) - \Pi_0(\bQ,0)$ is finite.

In the special case $q_t = 0$, the $k_0$ integral in Eq.~\eqref{Pi0krktint} is elementary. Setting $q_t = 0$, one obtains
\begin{subequations}
\begin{align}
 k_t^{++}(q_t=0) &= \left( \frac{2ik_0 - v_Fq_r}{2b} \right)^{1/4} \, , \\
 k_t^{--}(q_t=0) &= \left( \frac{2ik_0 - v_Fq_r}{2b} \right)^{1/4} i s(k_0) \, ,
\end{align}
\end{subequations}
where $s(k_0)$ is the sign of $k_0$. The residues simplify to
\begin{subequations}
\begin{align}
 R^{++}(q_t=0) &= \frac{1}{4} \left( \frac{2ik_0 - v_Fq_r}{2b} \right)^{-3/4} \, , \\
 R^{--}(q_t=0) &= \frac{i}{4} \left( \frac{2ik_0 - v_Fq_r}{2b} \right)^{-3/4} s(k_0) \, .
\end{align}
\end{subequations}
Inserting this into Eq.~\eqref{Pi0krktint} and introducing an ultraviolet cutoff $\Lambda$, we obtain
\begin{equation}
 \Pi_0(q_r,0,iq_0) = - \frac{1}{8bv_F} \int_{-\Lambda}^{\Lambda} \frac{dk_0}{2\pi} \,
 \Theta(|k_0| - |q_0|/2) \left( \frac{2ik_0 - v_F q_r}{2b} \right)^{-3/4}
 [1 + s(k_0)] \, .
\end{equation}
The frequency integration is obviously elementary. Subtracting $\Pi_0(0,0,0)$, we can take the limit $\Lambda \to \infty$, yielding
\begin{equation}
 \delta\Pi_0(q_r,0,iq_0) =
 \frac{1}{4\pi v_F (2b)^{1/4}} \left[
 (1-i) \sqrt[4]{i|q_0| - v_F q_r} + (1+i) \sqrt[4]{-i|q_0| - v_F q_r} \, \right] \, .
\end{equation}
Analytic continuation of this expression in the upper complex frequency half-plane to real frequencies yields Eq.~\eqref{dPi0qt0}.

For $q_t \neq 0$ we write $\Pi_0$ as a sum of two terms, $\Pi_0 = \Pi_0^+ + \Pi_0^-$, where $\Pi_0^+$ and $\Pi_0^-$ are obtained from the contributions with $k_0 > 0$ and $k_0 < 0$ to the integral in Eq.~\eqref{Pi0krktint}, respectively. Shifting the integration variable by $\pm |q_0|/2$, one obtains
\begin{eqnarray} \label{Pi0pm1}
 \Pi_0^-(\bq,iq_0) &=& - \frac{1}{4bv_F} \int_{-\infty}^0 \frac{dk_0}{2\pi} \,
 \frac{1}{\sqrt{2} \sqrt{q_t^4 + \frac{2ik_0 - i|q_0| - v_Fq_r}{b}}}
 \sum_{s=\pm1} \frac{1}{\sqrt{-\frac{3}{4} q_t^2 + \frac{s}{\sqrt{2}}
 \sqrt{q_t^4 + \frac{2ik_0 - i|q_0| - v_Fq_r}{b}}}} \, , \nonumber \\
 \Pi_0^+(\bq,iq_0) &=& - \frac{1}{4bv_F} \int_0^{\infty} \frac{dk_0}{2\pi} \,
 \frac{1}{\sqrt{2} \sqrt{q_t^4 + \frac{2ik_0 + i|q_0| - v_Fq_r}{b}}}
 \sum_{s=\pm1} \frac{1}{\sqrt{-\frac{3}{4} q_t^2 + \frac{s}{\sqrt{2}}
 \sqrt{q_t^4 + \frac{2ik_0 + i|q_0| - v_Fq_r}{b}}}} \, . \nonumber \\
\end{eqnarray}
The integrals in Eq.~\eqref{Pi0pm1} are UV divergent. Subtracting $\Pi_0^\pm(\bQ,0)$, one obtains finite expressions for
$\delta\Pi_0^\pm(\bq,iq_0) = \Pi_0^\pm(\bq,iq_0) - \Pi_0^\pm(\bQ,0)$.

Eq.~\eqref{Pi0pm1} can be continued analytically to the entire upper complex frequency half-plane by simply extending $i|q_0| \to z$ with $\Im z > 0$ (note that $|q_0| = q_0$ in the upper frequency plane). One can easily check that the integrands encounter no poles or branch cuts for $\Im z > 0$, for any $k_0$. Hence, the continuation to real frequencies is obtained by substituting $i|q_0| \to \om + i0^+$.
Moreover, since $k_0$ and $q_0$ enter via the linear combinations $2ik_0 - i|q_0|$ for $k_0 < 0$ and $2ik_0 + i|q_0|$ for $k_0 > 0$, the infinitesimal imaginary part $i0^+$ is redundant and can be dropped. At this point it is clear that $\delta\Pi_0^-$ and $\delta\Pi_0^+$ depend on $q_r$ and $\om$ only via the linear combinations $-\om - v_Fq_r$ and $\om - v_Fq_r$, respectively.
Substituting $k_0$ by $-k_0$ in the integral for $\Pi_0^-$, the analytic continuation of Eq.~\eqref{Pi0pm1} to real frequencies can be written as
\begin{equation} \label{Pi0pm2}
 \Pi_0^\pm(\bq,\om) = - \frac{1}{4bv_F} \int_0^{\infty} \frac{dk_0}{2\pi} \,
 \frac{1}{\sqrt{2} \sqrt{q_t^4 + \frac{\pm 2ik_0 \pm \om - v_Fq_r}{b}}}
 \sum_{s=\pm1} \frac{1}{\sqrt{-\frac{3}{4} q_t^2 + \frac{s}{\sqrt{2}}
 \sqrt{q_t^4 + \frac{\pm 2ik_0 \pm \om - v_Fq_r}{b}}}} \, .
\end{equation}
To obtain the scaling form Eq.~\eqref{dPi0}, we introduce a dimensionless integration variable $\tk_0$ defined by $k_0 = b|q_t|^4 \tk_0$, yielding
\begin{equation} \label{Pi0pm3}
 \Pi_0^\pm(\bq,\om) = - \frac{|q_t|}{4bv_F} \int_0^{\infty} \frac{d\tk_0}{2\pi} \,
 \frac{1}{\sqrt{2} \sqrt{1 \pm 2i\tk_0 + \frac{\pm \om - v_Fq_r}{bq_t^4}}}
 \sum_{s=\pm1} \frac{1}{\sqrt{-\frac{3}{4} + \frac{s}{\sqrt{2}}
 \sqrt{1 \pm 2i\tk_0 + \frac{\pm \om - v_Fq_r}{bq_t^4}}}} \, .
\end{equation}
Subtracting $\Pi_0^\pm(\bQ,0)$ with the same substitution we obtain 
\begin{equation}
 \delta\Pi_0^\pm(\bq,\om) =
 \frac{|q_t|}{4v_F} \, I^{\pm}\Big( \frac{\pm\om - v_Fq_r}{bq_t^4} \Big) \, ,
\end{equation}
with the dimensionless scaling functions
\begin{equation}
 I^\pm(x) = \int_0^\infty \frac{d\tk_0}{2\pi} \left[
 \frac{1 \pm i}{2 (\pm i\tk_0)^{3/4}} -
 \frac{\frac{1}{\sqrt{-\frac{3}{4} + \sqrt{ \frac{1}{2} (1+x \pm 2i\tk_0)}}} +
 \frac{1}{\sqrt{-\frac{3}{4} - \sqrt{ \frac{1}{2} (1+x \pm 2i\tk_0)}}}}
 {\sqrt{2(1+x \pm 2i\tk_0)}} \right] \, .
\end{equation}
Obviously $I^+(x)$ and $I^-(x)$ are related by complex conjugation, that is,
$I^+(x) = [I^-(x)]^*$. Denoting $I^+(x)$ as $I(x)$, we obtain Eqs.~\eqref{dPi0} and \eqref{Ix}.

\end{appendix}



\begin{thebibliography}{99}
\bibitem{fn_incomm} Commensurate wave vectors can be written as a linear
 combination of reciprocal lattice vectors with rational coefficients,
 while incommensurate wave vectors cannot. Obviously there are infinitly
 many possible rational coefficients, but the quantitative effects of
 commensurability decrease rapidly with the size of their denominators.
 In a one-loop calculation there is no difference between commensurate
 and incommensurate cases, as long as $2\bQ$ is not a reciprocal lattice
 vector.
\bibitem{bonetti} The data for $\chi_0(\bq,0)$ have been provided by Pietro Bonetti.
\end{thebibliography}
\end{document}